# Enhancement of the Seebeck effect in bilayer armchair graphene nanoribbons by tuning the electric fields

Thanh-Tra Vu[1], Thi-Kim-Quyen Nguyen[2], Thi-My-The Nguyen[2], Van-Cuong Nguyen[3], Van-Truong Tran[4*]

[1]Department of Physics, School of Education, Can Tho University, Can Tho, Vietnam
[2]School of Graduate, College of Natural Sciences, Can Tho University, Can Tho, Vietnam
[3]Department of Fundamental Science, University of Transport Technology, Hanoi, Vietnam
[4]EM2C, CentraleSupélec, Université Paris Saclay, CNRS, 92295 Châtenay Malabry, France
*vantruongtran.nanophys@gmail.com

**Abstract**

The Seebeck coefficient in single and bilayer graphene sheets has been observed to be modest due to the gapless characteristic of these structures. In this work, we demonstrate that this coefficient is significantly high in quasi-1D structures of bilayer armchair graphene nanoribbons (BL-AGNRs) thanks to the open gaps induced by the quantum confinement effect. We show that the Seebeck coefficient of BL-AGNRs is also classified into three groups $3p$, $3p + 1$, $3p + 2$ as the energy gap. And for the semiconducting BL-AGNR of width of 12 dimer lines, the Seebeck coefficient is found as high as 707 µV/K and it increases up to 857 µV/K under the impact of the vertical electric field. While in the semimetallic structure of width of 14 dimer lines, the Seebeck coefficient remarkably enhances 14 times from 40 µV/K to 555 µV/K. Moreover, it unveils an appealing result as the Seebeck coefficient always increases with the increase of the applied potential. Such BL-AGNRs appear to be very promising for the applications of the next generation of both electronic and thermoelectric devices applying electric gates.

## 1 Introduction

Graphene is a one-atom thick super material that exhibits amazing electronic, thermal and mechanical properties.[1–3] It is therefore a great candidate for many applications in different fields, particularly in electronic devices such as p–n junctions, transistors, and sensors [4–6] that have attracted so many researchers to pay attention during the last decade. However, it is difficult to achieve high on/off ratio of the current in graphene transistors as well as to switch states in other devices made of graphene due to the gapless characteristic of this material.[3,7]

The lack of a bandgap in graphene is also responsible for the weakness of the thermoelectric effect obtained for this material as the Seebeck coefficient (also referred to as the thermoelectric power) has been reported to be smaller than 80 µV/K.[8] Although the band structure can be modified by stacking two graphene layers to form bilayer structures, a bandgap is not observed in the undoped bilayer graphene.[9,10] However, it has been reported theoretically[11] and experimentally[12] that it can open a bandgap in bilayer graphene by applying a vertical electric field. As the Seebeck effect is associated directly with the size of the bandgap,[13] it is thus possible to control this effect in bilayer structures by using external electric fields.

Recently, the electronic and thermoelectric properties of bilayer graphene nanoribbons subjected to external fields have been investigated.[14–19] It has shown that the bandgap in bilayer ribbon structures is strongly modulated under the impact of electric fields.[18,19] In bilayer zigzag graphene nanoribbons (BL-ZGNRs),[18] an energy gap up to 600 meV has been achieved in narrow ribbon structures and under the simultaneous influence of vertical and transverse electric fields. This bandgap induced by the presence of electric fields in BL-ZGNRs is much higher than that obtained by the same effect in 2D bilayer graphene sheets which is about 250 meV.[11,20] The Seebeck coefficient in BL-ZGNRs is also predicted to be as high as 700 µV/K corresponding to the largest bandgap modulated by electric fields.[18]

The impact of electric fields on the electronic properties of bilayer ribbons with armchair edges (BL-AGNRs) has been observed differently from what is found in BL-ZGNRs as the largest gap is achieved by solely the vertical electric field.[19] Although previous works[16,19,21,22] have explored the modulation of the band structure and the bandgap under the effect of vertical electric fields on BL-AGNRs, the transport properties have been not considered to fulfill the scattering picture of electrons in this kind of structures.

In this work, we investigate the electronic and thermoelectric properties of BL-AGNRs subjected to a vertical electric field that perpendicular to the graphene sheets. By means of simulations with Tight Binding (TB) calculations and Green's function formalism, we demonstrate that the bandgap of BL-AGNRs can be altered in a wide range by tuning the electric field. Additionally, the study of the transport properties depicts that the intrinsic Seebeck coefficient of BL-AGNRs without electronic fields also classifies into three groups $3p$, $3p + 1$, $3p + 2$ with $p$ is an integer number, similar to that obtained for the bandgap in single layer AGNRs[23,24] and BL-AGNRs.[19] More interestingly, although the bandgap fluctuates with the increase of the applied voltage, it shows that the Seebeck coefficient substantially grows up. These results not only unfold the electronic and thermoelectric transport properties of BL-AGNRs under the impact of external fields but also provide a more comprehensive understanding of the relationship between the Seebeck coefficient and the energy and transport gaps.

## 2 Modeling and Methodologies
### *2.1 Modeling*

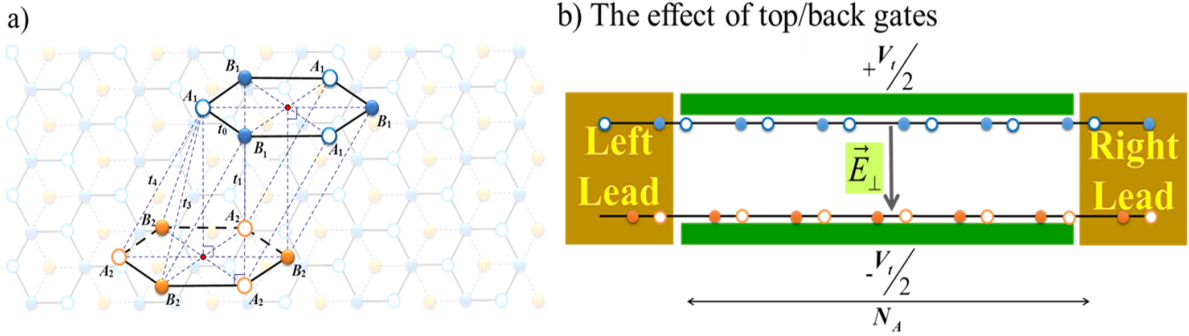

**Figure 1**. a) Illustration of a BL-AGNR structure, the intra-layer and inter-layer interactions between atoms in a single layer and two layers. b) the BL-AGNR under the effect of a vertical field generated by a top gate $+V_t/2$ and a bottom gate $-V_t/2$.

The studied structure is sketched as in Fig. 1(a) in which the blurred shape illustrates a bilayer graphene with armchair edges and the emerged one describes the interactions between atoms in a single layer or in different layers. As the two layers are coherent by the Bernal stacking, $A_2$ sites of the lower layer (orange) are located exactly at the position of $B_1$ sites of the upper layer (blue). The length of C and C bond is 0.142 nm and the distance between the two layers is 0.335 nm. The width of the BL-AGNR is characterized by number of dimer lines $M$ along the width of each sub-ribbon.[19]

The parameters $\{t_0, t_1, t_3, t_4\}$ are the hoping energies in which $t_0$ is the intra-layer nearest-neighbor interaction between carbon atoms at A and B sites while $t_1$, $t_3$ and $t_4$ present the inter-layer interactions between the nearest atoms (sites A-B), the next-nearest atoms (sites A-B) and the next-nearest atoms (sites A-A, B-B) in two layers, respectively. [25–27]

To explore the effect of the vertical electric field on the electronic structure and the transport properties of BL-AGNRs, a bottom and top gates with electrostatic potentials $+V_t/2$ and $-V_t/2$, respectively, were added as sketched in figure 1(b).

## 2.2 *Methodologies*

To examine the electronic and thermoelectric properties of GL-AGNRs, a Tight Binding model was utilized. The Hamiltonian of the system subjected to an external electric field is written as [28]

$$H = \sum_i U_i |i\rangle\langle i| - \sum_{\langle i,j \rangle} t_{ij} |i\rangle\langle j|, \quad (1)$$

where $U_i$ is the total potential energy at *i*-th site, $U_i = -eV_t/2$ if the *i*-th site belongs to the upper layer and $U_i = +eV_t/2$ if this site belongs to the lower layer. $t_{ij}$ is the hopping energy between the atom at *i*-th site and its surrounding neighbor atoms and $t_{ij}$ will be fitted to $t_0$, $t_1$, $t_3$ or $t_4$ depending on the distance between the *i*-th and *j*-th atoms. In our calculations, $t_0 = 2.598$ eV, $t_1 = 0.364$ eV, $t_3 = 0.319$ eV and $t_4 = 0.177$ eV were taken from refs. [14,25] where they were deduced by a fitting to *ab initio* results.

To study transport properties, we used the non-equilibrium Green's function (NEGF) approach within the ballistic approximation [29–31]. The electrical conductance and the Seebeck coefficient were calculated as [32,33]

$$\begin{cases} G_e(\mu,T) = e^2 L_0(\mu,T) \\ S(\mu,T) = \dfrac{1}{eT} \dfrac{L_1(\mu,T)}{L_0(\mu,T)} \end{cases} \quad (2)$$

where $L_n(\mu,T)$ is the intermediate function and determined by equation [32]

$$L_n(\mu,T) = \frac{2}{h} \int_{-\infty}^{+\infty} dE\, T_e(E)(E-\mu)^n \frac{-\partial f_e(E,\mu,T)}{\partial E}, \quad (3)$$

in which $T_e(E)$ is the electron transmission defined within Green's function formalism,[30] $T_e(E) = Trace(\Gamma_L G \Gamma_R G)$, here $\Gamma_{L(R)} = i(\Sigma_{L(R)} - \Sigma_{R(L)})$ is the injection rate at the interfaces of the left (right) leads and the active region, and $G = [E - H_D - \Sigma_L - \Sigma_R]^{-1}$ is the retarded Green's function of the central region. Also in equation (3), $f_e(E,\mu,T)$ is the Fermi distribution function, $\mu$ is the electron chemical potential and $T$ is the absolute temperature.

# 3 Results and discussions

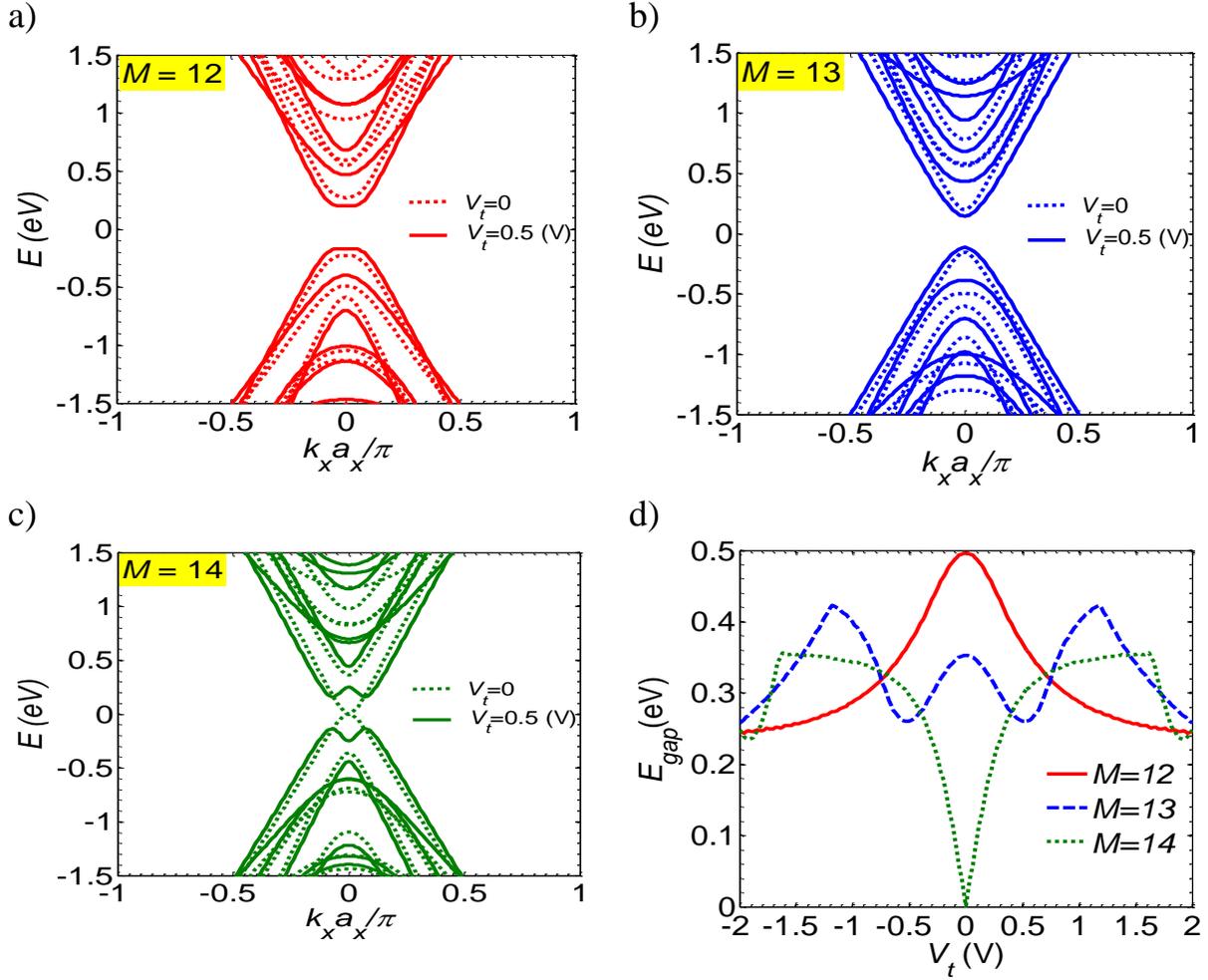

**Figure 2**. The band structure without and with the electric field applied for the three considered structures (a) $M = 12$, (b) $M = 13$, and (c) $M = 14$, respectively. (d) The energy gap is plotted as a function of $V_t$.

## 3.1 The impact of the vertical electric field on the band structure of BL-AGNRs

We first consider the band structure of BL-AGNRs without and with electric fields for the structures of width $M = 12$, 13 and 14 which characterize for the three typical groups $M = 3p$, $M = 3p+1$ and $M = 3p+2$, respectively (with $p$ as an integer number). In Figs. 2(a), 2(b) and 2(c), the dotted lines are results in the absence of the vertical electric field while the solid ones present the outcome in the presence of the electric field for an applied bias $V_t = 0.5$ (eV). As it can be observed from the dotted lines in these panels, the energy gap is found in the structures of width $M = 12$ (group $3p$) and $M = 13$

(group 3p + 1) which thus demonstrate these families are semiconducting. In contrast, without external fields applied, the structure $M = 14$ (group 3p + 2) exhibits as semi-metallic and in agreement with the conclusions in previous studies.[19]

In the presence of a vertical electric field, the bandgap of former groups becomes narrower whereas it increases in the later one.

To further investigate the dependence of the bandgap on the strength of the vertical electric field, we dissect the bandgaps of all these three groups as a function of the potential $V_t$ as shown in Fig. 2(d). Observing the evolution of the bandgap in all structures, it can be seen that the gap is remarkably modulated even with small biases in the range $V_t = [0, 0.5\ V]$, i.e., the gap is suppressed from 0.5 eV to 0.36 eV and from 0.35 eV to 0.26 eV in the structures $M = 12$ and $M = 13$, respectively. In an opposite trend, the gap in the structure $M = 14$ is enhanced from zero to 0.3 eV. These results with the reduction of the bandgap in semiconducting structures and the open of the gap in semi-metallic ones are similar to that obtained in single layer graphene nanoribbons.[34,35] However, an interesting result is observed from the dashed blue line of the structure $M = 13$ as the bandgap turns back when the potential is larger than 0.5 V. The bandgap of this structure reaches the maximum value max($E_{gap}$) = *0.42 eV* at the applied potential $V_t = \pm 1.2$ V and it is thus larger than the one without the vertical electric field applied (0.36 eV). This outcome can be considered as a peculiar feature of BL-AGNRs under the effect of the vertical electric field and it might find potential applications in relevant fields.

### 3.2 *The impact of the vertical electric field on the Seebeck effect*

In this section, the Seebeck effect without and with electric fields applied will be considered.

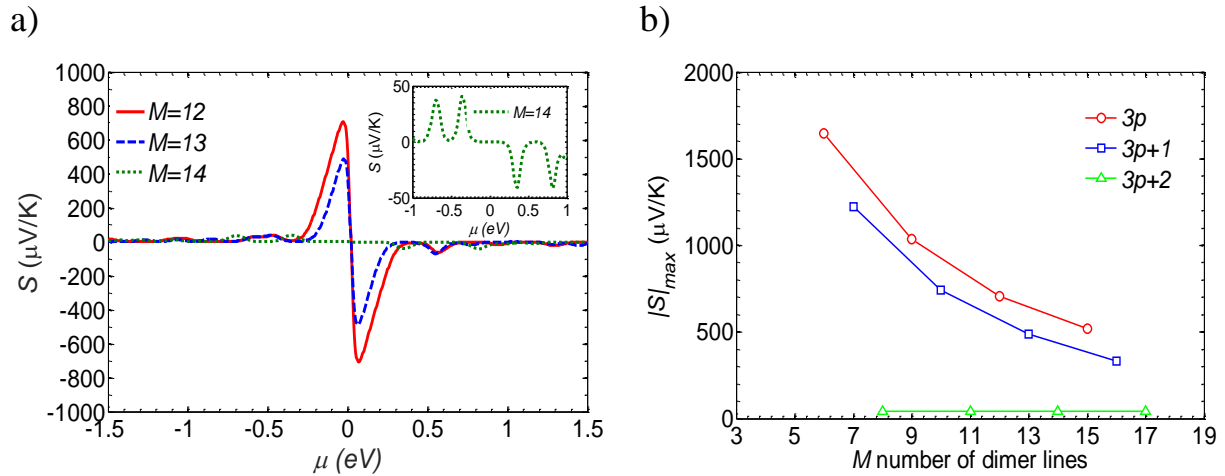

**Figure 3**. a) The Seebeck coefficient is plotted as a function of the chemical energy for the three studied structures. b) The maximum Seebeck coefficient is considered as a function of number of dimer lines *M* for three typical groups $3p$, $3p+1$, and $3p+2$. The results are obtained at the room temperature and for the case without the electric field applied.

### 3.2.1  The intrinsic thermoelectric properties of BL-AGNRs

We first examine the intrinsic thermoelectric ability of the undoped BL-AGNRs. In Fig. 3 (a) the thermoelectric power at the room temperature is shown as a function of the electron chemical energy $\mu$. As it can be seen clearly in this figure, the maximum value of the Seebeck coefficient and the position of the peaks are strongly dependent on the width of the structures *M* = 12, 13 and 14 as shown correspondingly by three lines: the solid red, dashed blue and dotted green lines, respectively, for a comparison. The Seebeck coefficient exhibits two highest peaks that almost symmetrically via the zero energy point and lie in the range [-0.5 eV, 0.5 eV] of the chemical energy. Here the positive and negative peaks depict the contribution of the positive (holes) and negative (electrons) carriers to the Seebeck coefficient, respectively. The red line displays the highest peak with the maximum Seebeck coefficient $|S|_{max} = 707\,\mu V/K$ while the dashed blue and dotted green (in the inset) have $|S|_{max} = 487$ µV/K and 40 µV/K, respectively. The obtained results are consistent with the correlation of the size of the bandgaps in these structures that shown in Fig 2.

As it has been demonstrated that the bandgap in BL-AGNRs and AGNRs strongly depends on the width of the structures,[19,23] it is thus relevant to investigate the interplay of the Seebeck coefficient and the number of dimer lines *M*. The maximum Seebeck coefficient of BL-AGNRs was deduced as a function of *M* for three characteristic groups of the width $3p$, $3p + 1$ and $3p + 2$ and shown in Fig. 3(b). It can be observed clearly that the Seebeck coefficient is also classified into three families and in the same order as the energy gap[19] in which $\left(|S|_{max}\right)_{3p} > \left(|S|_{max}\right)_{3p+!} > \left(|S|_{max}\right)_{3p+2}$. Basically, $|S|_{max}$ reduces with *M* in all groups and thus suggesting that higher Seebeck coefficient is found in narrower structures. It is worth to note that the Seebeck coefficient of the group $3p + 2$ stays constantly at a small value ($\approx 40$ µV/K) and it is consistent with the fact that the bandgap of this group is found to be very small or almost equal zero.[19]

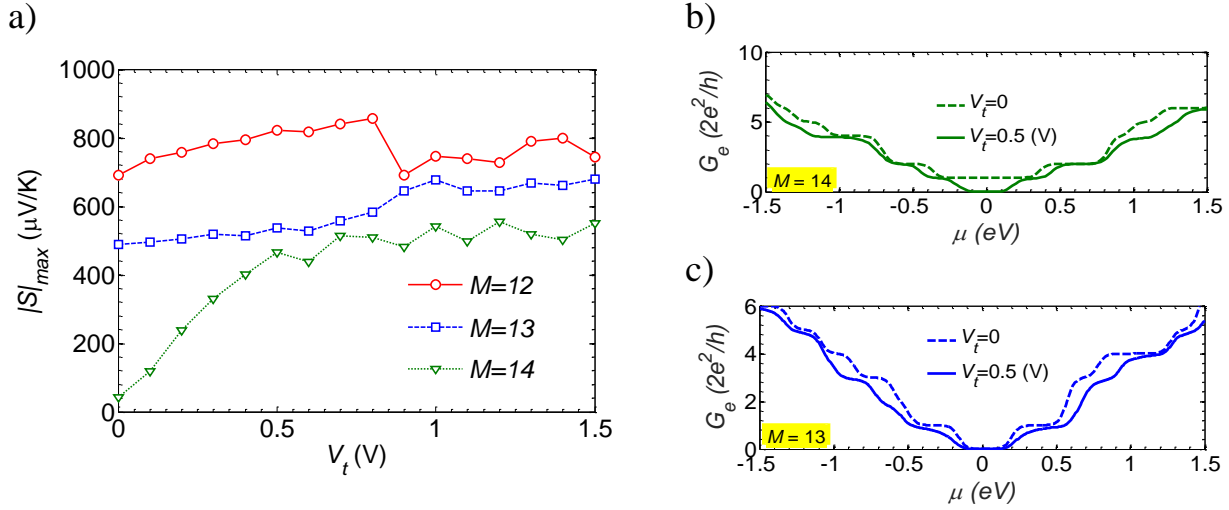

**Figure 4**. a) The evolution of the maximum Seebeck coefficient with the increase of the potential $V_t$. The electrical conductance without and with the electric field $V_t = 0.5$ (V) applied to the structures (b) $M = 14$ and (c) $M = 13$. The simulations were performed at 300 K.

### 3.2.2 Modulation of the Seebeck coefficient by tuning electric fields

As it has been demonstrated, the Seebeck coefficient is associated directly to the bandgap, it is thus possible to control this parameter by tuning the bandgap. Here, we will examine the impact of the vertical electric field on the Seebeck effect in BL-AGNRs. As it is usually utilized in the transport study, the electric field is applied only to the central region (Fig. 1(b)) which is connected with two semi-infinite leads made of undoped BL-AGNRs. The maximum Seebeck coefficient $|S|_{max}$ was then calculated as a function of $V_t$ for the three structures $M = 12$, 13 and 14 with the same length of the central region $N_A = 50$ unit cells ($\approx 21$ nm). Fig. 4(a) displays the results at the room temperature.

First, the dotted green line with triangular symbols shows a remarkable enhancement of $|S|_{max}$ in the structure of width $M = 14$. Particularly, in the range of the potential $V_t = [0, 0.5\ V]$, $|S|_{max}$ increases sharply from 40 µV/K to 466 µV/K, then it grows weakly with a slight fluctuation and the maximum value can be obtained is about 555 µV/K, which is thus about 14 times higher than that without the electric field applied. This strong enhancement of the Seebeck coefficient in the structure $M = 14$ is understandable owing to the bandgap is open widely as observed in Fig. 2(d). However, in the same range of the bias potential, $|S|_{max}$ of the other groups is also observed to be enhanced while the bandgap is demonstrated to be reduced as shown in Fig. 2(d). This result thus presents an abnormal relationship of the Seebeck coefficient with the bandgap as they usually vary in the same way.

To understand this result, we calculated the electrical conductance (at 300 K) and the results are shown in Figs. 4(b) and 4(c). It should be noted that the valley region where the electrical conductance is almost equal zero is often addressed as the transport gap or the conductance gap and it might be different from the energy gaps due to the scattering occurred in the central region. As a transport quantity, the Seebeck coefficient should behave similarly to the electrical conductance and directly depends on the transport gap. In Fig. 4(b) we display the result obtained for the structure $M = 14$. As it can be seen, the gap appears in the solid green line when the vertical electric field is applied and it thus explains for the strong enhancement of the Seebeck coefficient obtained in Fig. 4(a). In Fig. 4(c), the conductance without and with the electric field applied is shown for the structure of width $M = 13$. A similar result is found for the case $M = 12$. Here we can see that the conductance gap is wider under the impact of the electric field and thus well interpret the increase of $|S|_{max}$ in this structure. In fact, the inhomogeneity of the central region and the two leads when an electric field applied is the origin of the scattering in the central region which leads to a degradation of the electrical conductance around the Fermi energy. The decrease of $G_e$ results in a wider transport gap and a stronger separation of electrons and holes, it must be thus a larger Seebeck coefficient. This result provides a more comprehensive understanding of the relationship of the Seebeck effect and the energy and transport gaps.

Additionally, Fig. 4(a) depicts that with bias potential $V_t > 0.5$ V, $|S|_{max}$ reaches to the maximum values 857 µV/K at $V_t = 0.8$ V and 677 µV/K at $V_t = 1.0$ V in the structure $M = 12$ and 13, respectively.

The fluctuation of $|S|_{max}$ at high potentials in Fig. 4(a) can be understood as the bottom of the lowest conduction bands and the top of the valence bands are distorted around the Gamma points as observed in Figs. 2(a), 2(b) and 2(c) which can lead to a complex behavior of the scattering phenomenon.

### 3.2.3 Dependence of the Seebeck coefficient on the temperature

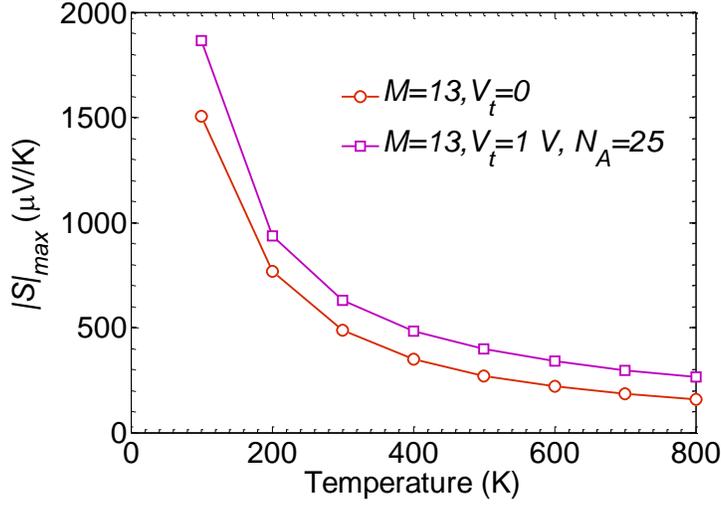

**Figure 5**. The dependence of the Seebeck coefficient on the temperature in the BL-AGNR of width $M = 13$ for both cases without and with the vertical electric field.

To complete the investigation of the Seebeck effect in BL-AGNRs, we also examine the variation in the Seebeck coefficient as the temperature changes for both cases without and with the electric field. The result is displayed for only the structure of width $M = 13$ while the similar outcome is observed for other structures of width. First, $|S|_{max}$ is clearly enhanced under the effect of the electric field at all temperatures, confirming again the usefulness of the vertical field. Second, $|S|_{max}$ is strongly reduced with the increase of the temperature and tends to saturate at very high temperatures, i.e., in the absence of the electric field (red line with circles), $|S|_{max}$ drops remarkably from 1500 µV/K at 100 K to 487 µV/K at 300 K. The reduction is slower at higher temperatures as $|S|_{max}$ is found equal 270 µV/K at 500 K. This inverse proportion of the Seebeck coefficient and the temperature is, in fact, consistent with what has been observed in previous works.[36]

## 4 Conclusions

In summary, we have investigated the band structure, the electrical and thermal properties of BL-AGNRs in the absence or the presence of the vertical electric field applied. By means of simulation, we have demonstrated that the energy gap of BL-AGNRs is strongly modulated by the external field. Notably, the gap is open remarkably in the structure $M = 14$ as it increases from almost zero to larger than 0.3 eV just by an applied bias of about 0.5 V. The transport study has unveiled that in the case without electric field applied, the Seebeck coefficient is also classified in to three groups as what has been obtained for the energy gap and the amplitude of this coefficient is proportional to the size of the energy gap as at the room temperature, $|S|_{max}$ = 707 µV/K, 487 µV/K and 40 µV/K corresponding to the gaps equal 0.5 eV, 0.35 eV and 0.0 eV in the structures $M = 12, 13$ and $14$, respectively. However, in the presence of the vertical electric fields, an intriguing result has been obtained in which the Seebeck coefficient does not vary in the same way as the energy gap and it always enhances for all structures. The Seebeck coefficient has been found to increase up to 857 µV/K, 677 µV/K and 555 µV/K in the structures $M = 12, 13$ and $14$, respectively just by utilizing moderate electric fields. This study not only fulfills the scattering picture in BL-AGNRs under the effect of external electric fields but also provides a more comprehensive understanding of the dependence of the Seebeck coefficient on the energy and transport gaps.

## Acknowledgments

This research is funded by Vietnam National Foundation for Science and Technology Development (NAFOSTED) under grant No. 103.01-2015.98.